# Analysing health professionals' learning interactions in online social networks: A social network analysis approach

XIN LI[a, b,1], KATHLEEN GRAY[a, b], KARIN VERSPOOR[a, b], STEPHEN BARNETT[c]
[a] *Health and Biomedical Informatics Research Centre*
[b] *Department of Computing and Information Systems, School of Engineering, University of Melbourne*
[c] *General Practice Academic Unit, Graduate School of Medicine, University of Wollongong, Australia*

**Abstract**

*Online Social Networking may be a way to support health professionals' need for continuous learning through interaction with peers and experts. Understanding and evaluating such learning is important but difficult, and Social Network Analysis (SNA) offers a solution. This paper demonstrates how SNA can be used to study levels of participation as well as the patterns of interactions that take place among health professionals in a large online professional learning network. Our analysis has shown that their learning network is highly centralised and loosely connected. The level of participation is low in general, and most interactions are structured around a small set of users consisting of moderators and core members. The structural patterns of interaction indicates there is a chance of small group learning occurring and requires further investigation to identify those potential learning groups. This first stage of analysis, to be followed by longitudinal study of the dynamics of interaction and complemented by content analysis of their discussion, may contribute to greater sophistication in the analysis and utilisation of new environments for health professional learning.*

## 1. Introduction

As medical knowledge is expanding and health care delivery is becoming more complex, health professionals must commit to continuous learning to maintain up-to-date knowledge and skills. One approach claimed to meet their learning and development needs is through the use of Online Social Networks (OSN) [1]. OSN has been found useful to reduce professional isolation and support anytime-anywhere peer-to-peer interaction at scale [2, 3]. Also, it contributes to the development of professional networks and improves Continuing Professional Development (CPD) [4].

However, there is a lack of understanding about how learning occurs in OSN, making it difficult to design and facilitate this type of learning. There are acknowledged challenges in developing effective online discussions for health professionals [5]. To realise the full potential of OSN for health professionals' learning, the Institute of Medicine [6] suggests evaluating this type of learning and integrating effective approaches into the mainstream of CPD.

Although understanding and evaluating this type of learning is important, traditional evaluation approaches are unsuited to the task. They may focus more on learning outcomes, whereas the learning occurring in OSN has greater emphasis on the sharing of opinions and the validation of current practice, and therefore the focus of evaluation is more on learning processes [7]. To study health professionals' learning in their OSN, the first step is to identify and understand the interactions within their learning environment, as this helps understand their learning behaviours which provides valuable insights into learning results [4].

Social Network Analysis (SNA) has already proven to be an effective technique to analyse interaction in online learning environments [8, 9]. Using SNA, the nodes (learners, or learning resources) and ties (relationships) in networks can be visualised and analysed using quantitative measures and graphical representations in order to examine the flow of interactions. The aim of this paper is to demonstrate how SNA can be used to study levels of participation as well as the patterns of interactions that take place among health professionals in a large online professional learning network.

## 2. Background and Related Work

Recent technological changes, in particular network technologies such as OSN, have reorganised how we learn and brought us *networked learning*. Goodyear [10] defines networked learning as the learning in which information and communications technology is used to promote connections between learners, and between a learning network and its learning resources. The study of networked learning aims to understand the learning process by investigating how people develop and maintain a web of social relations for their learning; it focuses on the diversity of social relationships (rather than the development of long-lasting relationships), as well as the value this creates for learning.

---
[1] Corresponding Author.

Based on the theory of social constructivism, learning best occurs through social interactions and communication between learners [11]. The learning theory called *connectivism* states that learning resides in networks. These networks are formed from the social interactions between learners. The theory emphasises the importance of finding new connections; the learning process begins with establishing and finding new connections [12].

The application of SNA to learning is still at a very early stage [13]. In particular, there have been very few studies conducted in the context of informal learning for professional development in any specific field. SNA was employed to capture and analyse traces of teachers' informal learning occurring in their social-professional networks [14]. SNA was also used to understand the flow of experiential knowledge sharing among health professionals within a paediatric pain discussion forum [15]. However, both studies were limited in their analysis, and based on small numbers.

## 3. Methods

### 3.1. Data

Data for this study comes from an online professional learning network used by more than 10,000 health professionals during the period 2009-2014. The online learning network is accessed by health professionals only. All doctors were verified by entering their Medical Registration details, which were then checked against the registration database before users gained entry. The online community has forums set up which allow the doctors discussing industry issues, sharing best practices and promoting conversation within the health community. Topics of discussion tend to be highly clinically focused, for example, cardiovascular, rheumatology, dermatology, etc.

### 3.2. Measures

The interactions of health professionals within the forum were analysed using different SNA measures, which includes both mathematical and visual approaches. As Table 1 depicts, mathematical analysis involves using four network-level structural measures (density, centralisation, diameter and average path length) and three individual-level centrality measures (degree, betweenness, and closeness). The network structural measures reveal the participation level and connectivity in the entire network. The centrality measures provide information about the activity levels of the individual users, along with the overall activity status of the network; they help understand how interactions take place by summarising the individual users/thread-level characteristics. Lastly, visual analysis was done by representing the relationships between users/threads through graphs to enrich the findings of mathematical analysis.

**Table 1 – Descriptive Definitions of All SNA Measures Used in This Study**

| SNA measure | | Descriptive definition |
|---|---|---|
| Network structural measures | Density | The number of present relationships as a ratio of the possible number of relationships in a network; represents the overall connection between users/threads. |
| | Centralisation | The extent to which the connectedness is focused around a particular user/thread. |
| | Diameter | The longest path between any pair of users/threads in a network. |
| | Average path length | The average path between any pair of users/threads in a network. |
| Centrality measures | Degree centrality | The number of direct relationships a user/thread has with others in a network, which provides an indication of their popularity and influence. |
| | Betweenness centrality | The number of times a user/thread sits on the shortest path linking two other users/threads together in a network; it helps identify important users/threads. |
| | Closeness centrality | How quickly a user/thread can reach all other users/threads within the entire network; it provides an indication of the speed of information distribution. |

### 3.3. Procedure

An online discussion forum is a type of 2-mode network, which represents how actors are tied to particular events (i.e. an actor-by-event network). A common method of analysing a 2-mode network is to transform the data into two 1-mode networks [16]. One is an actor network, in this case, created from the forum users. A tie is created between them if they both communicate on the same thread. Another is an event network, created from discussion threads. A tie is created between two threads if the same user communicated on both of them.

According to this, once the forum's data is extracted through SQL queries, it is structured into two matrices for analysis. The statnet library in R was used for the analysis and visualisation of interactions. Firstly, the network structural measures were applied to both user (actor) and thread (event) networks, and then centrality analysis was performed for both of the networks. The results of density, centralisation, and centrality analysis were normalised for the adaptation to 2-mode networks [17] and to a [0, 1] scale for simpler interpretation.

## 4. Results and Discussion

### 4.1. Network Activities Overview

In order to analyse the online professional learning network for health professionals, basic statistics were extracted first to provide basic understanding of activity levels within the network. There are a total of 10056 registered users in the forum, mainly with backgrounds of General Practice (n=7632), but also including Nursing (n=775), Cardiology (n=125), and General Medicine (n=122).

In the study period 2009-2014, there were 40 forums, with total of 723 threads and 7089 posts. 621 users posted at least once. Of these 621, most users made less than 100 posts; five made more than 200 posts, however, three of them are known to have formal roles as moderators. When the post count for each forum was analysed, we found post events not only varied over time but were different across different forums. Politics/IT/administration, Doctors' life, and Cardiovascular/vascular were the most popular forums. A number of topics were discussed in 2011 but stopped after 2012. Discussion on specialised topics (e.g. Paediatric, Neurology) was not very intense but periodic, in particular, in the year 2012 and 2014.

### 4.2. Network Structural Measures

Table 2 presents the network structural measures for both user and thread networks. With regard to the user network, the density score of 0.04 indicates a low level of participation and connection among the users in the network. The centralisation score of 0.59 indicates that the interaction is centralised in a small set of users and a great amount of users are not engaged and interact as little as possible. A diameter of 5.00 and average path length of 2.17 indicates that the users are not very close to one another, which confirms the low density of the network, meaning that the users may not easily reach each other and share knowledge.

The thread network has a much higher density of 0.40 compared to the user network, which may imply that the same users initiated a large number of threads, and that those threads were connected through them. The centralisation score of 0.46 indicates that most threads were attended equally and only a few attracted more attention than others. The diameter of 4.00 and average path length of 1.66 indicate that the threads sit close to each other in general, however some are distant from each other. This implies some threads were initiated and commented by different set of users and so there is a chance of small group learning occurring.

**Table 2 – Network Structural Measures**

| Network structural measure | User (N=621) | Thread (N=723) |
|---|---|---|
| Density | 0.04 | 0.40 |
| Centralisation | 0.59 | 0.46 |
| Diameter | 5.00 | 4.00 |
| Average path length | 2.17 | 1.66 |

### 4.3. Centrality Measures

Figure 2 presents the centrality distributions for the users. As shown, degree distribution is quite skewed, ranging from 0.001 to 0.42 with a median of 0.003. It is a highly centralised network, with a minority of users who have degree of 0.20 or above producing the bulk of the discussion with the network. These users were found to include moderators and core members. Moderators are those that initiate and mostly encourage discussions, core members are self-directed learners who initiate and participate most discussions.

Closeness centrality ranges from 0.26 to 0.59 with a median of 0.37. Most of the users are not very close to each other, confirming that the network is loose. However, most users are similarly positioned, which implies that the users have potential access to learn from each other.

Betweenness centrality has a highly skewed distribution. There are a large number of users who have 0.00 betweenness. Only a few users have betweenness centrality of 0.02 or above. Median betweenness centrality is 4.83886e-06 which is very low, showing that most of the users are a part of a cluster; the cluster forms the core of the network and are not well connected each other. This low connectivity further emphasises the very weak participation in the forums.

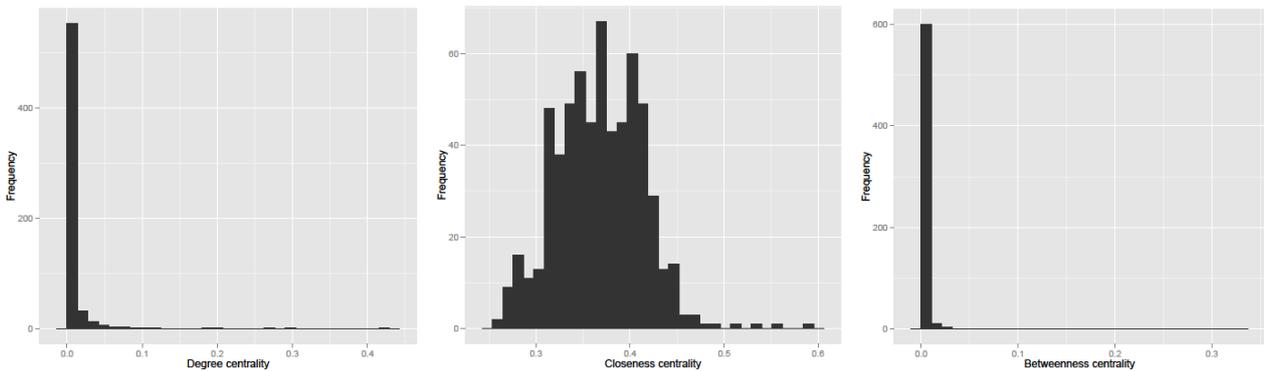

**Figure 2 – The Centrality Distributions for the Users**

Figure 3 depicts the centrality distributions for the threads. Degree distribution ranges from 0.002 to 0.05 with a median of 0.006, indicating there is no one thread that everyone commented on. The thread network is not centralised compared to the user network. There are only a few threads with relatively higher degrees of 0.024 or above, as the majority of threads had fewer than 13 comments.

Closeness centrality ranges from 0.43 to 0.59 with a median of 0.45, indicating that most of the threads sit close to each other, and are similarly positioned. This suggests that users who participated in one learning topic did not find it difficult to participate in other learning topics.

Betweenness centrality ranges from 0.00 to 0.02 with a median of 0.001. Though a large number of threads have 0.00 betweenness, this number is much lower than in the user network. More than 20% of the threads have betweenness centrality of 0.003 or above, showing that the threads are independent, and there is no one particular thread required to initiate other threads, or to keep other threads going.

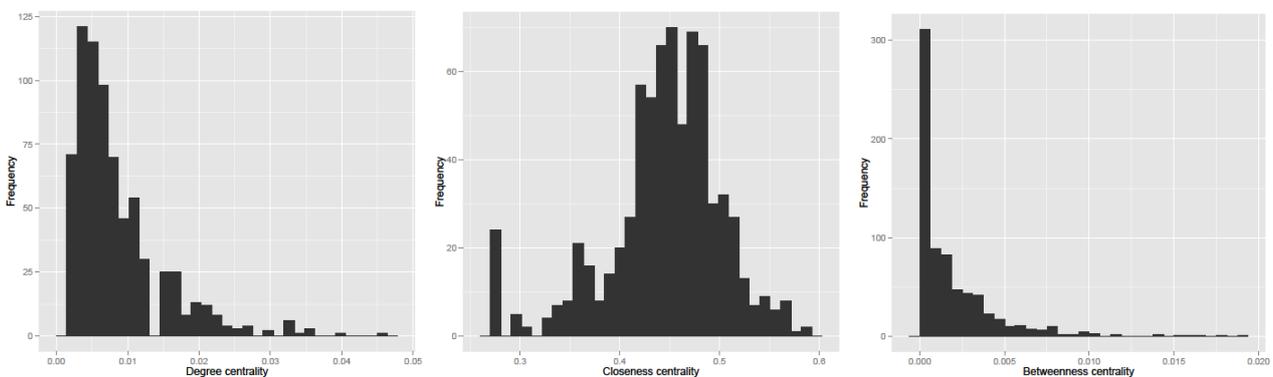

**Figure 3 – The Centrality Distributions for the Threads**

### 4.4. Network Visualisation

Figure 4 presents the visualisation of the 2-mode network, which has nodes of user and thread. We optimised the layout by applying the layout algorithm that directs most connected nodes into the centre of the graph. As shown, there are only a small number of users sitting in the centre of threads who were really engaged and actively participating in threads: about seven active users controlled most threads where comments were posted. Interestingly, there was one user (who is not in the central) initiated more than 20 threads but had no responses.

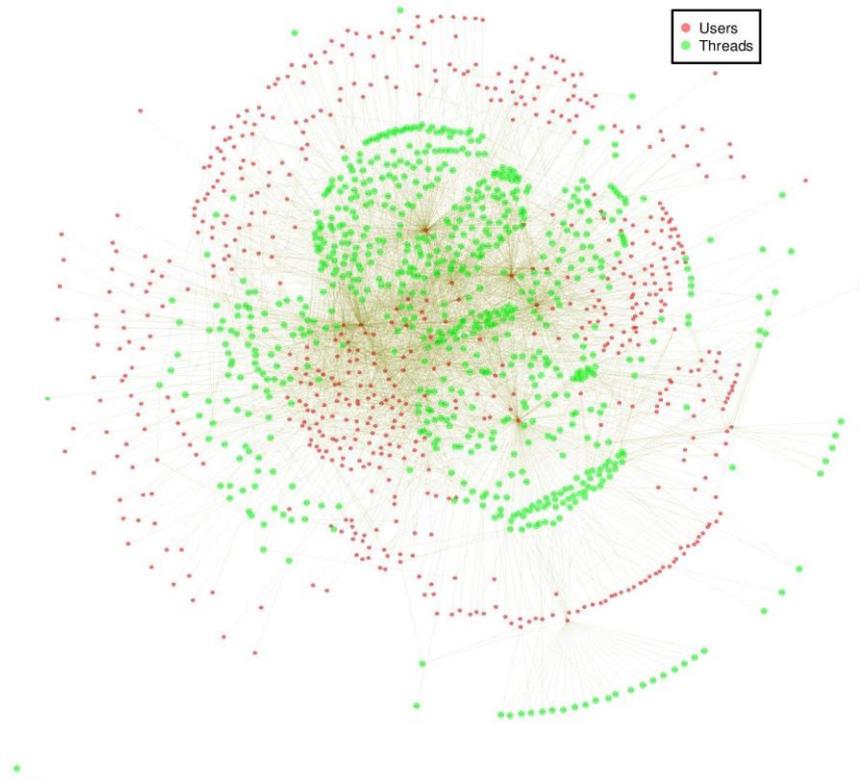

**Figure 4 – 2-mode Network**

Figure 5 presents the visualisation of user and thread networks separately. Given the large size of the network (users=621, threads=723), we thinned both networks by displaying only those ties that satisfy a cut-off point (specifically, we kept only those ties that have a tie weight greater than the mean tie weight plus one standard deviation). As shown, the user network (node size represents the number of threads that the user contributed) is quite centralised. The conversation mostly occurs among a few active users and the rest are not engaged much. The thread network is generally decentralised. Some threads received more attention than others, but were quite independent of each other. All of these findings are confirmed by the network structural and centrality analysis results.

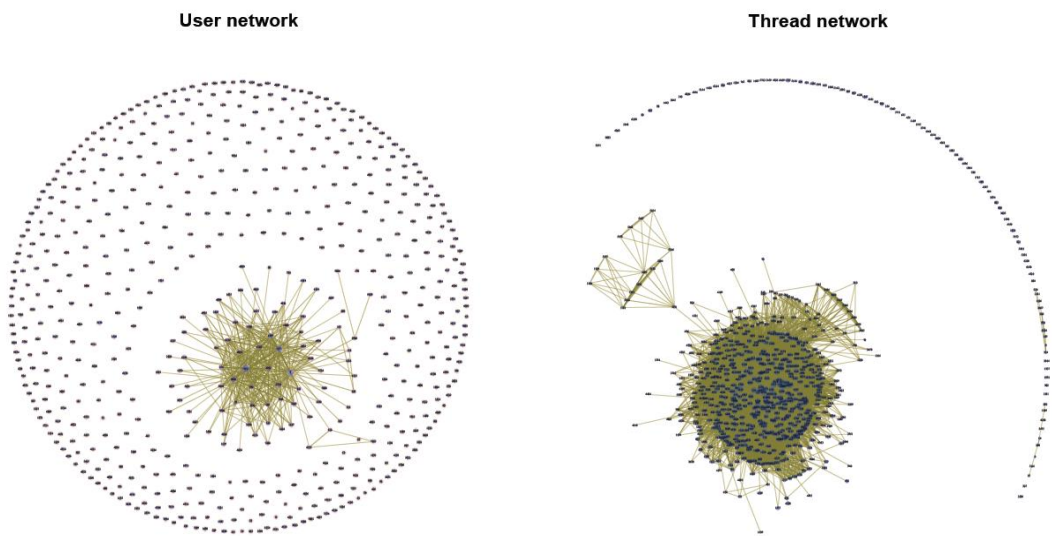

**Figure 5 – 1-Mode Network**

## 5. Conclusions and Future Work

OSN has potential as an innovative approach to informal learning for professional development of health professionals. However, we need to gain a clear understanding of how the process of online interaction can be considered to be educational. This study takes the first step by analysing the interactions occurring in a large online professional learning network for health professionals using SNA.

We transformed the online learning network (a discussion forum) into a user and thread network, and analysed the interactions within two networks respectively using network-level structural and individual-level centrality measures. We concluded that the participation level is low in general and the learning network is highly centralised and loosely connected. This finding is consistent with other research [18, 19] that has found evidence of a small set of users producing the bulk of the discussion within online communities. In our study, the small set of users consists of moderators and core members. Since they control what knowledge and information is shared, engaging them is essential to furthering the interaction or learning process in the network. In addition, the structural patterns of interaction indicate that there is a chance of small group learning occurring; this requires further investigation to identify potential learning groups.

Due to limitations of the data source, we were unable to track passive users (i.e. those who read but do not participate in any discussion) who are likely to gain value from the forum and may need support on their learning activities. Future studies should investigate their interactions with learning resources if such data is obtainable.

This study provides an indication of the participation level and the patterns of interactions within the forum; however, it does not necessarily inform to what extent these interactions impact learning. Therefore, it remains difficult to interpret learning based on the patterns of interactions. Longitudinal network analysis has been suggested to overcome this challenge by studying the interaction changes over time and the driving factors behind such changes [20]. Complemented by content analysis of the discussions in the forum, such analysis may help to explain how health professionals' knowledge is constructed and influenced by their interactions.

Studies such as these are expected to contribute to greater sophistication in the analysis of new environments for health professional learning, and thus to more effective design and operation of such learning environments.